# PROTECTION OF THE 6 T YBCO INSERT IN THE 13 T NB3SN FRESCA II DIPOLE

A. Stenvall, E. Haro, Tampere University of Technology, Department of Electrical Engineering, Electromagnetics, Tampere, Finland
Ph. Fazilleau, M. Devaux, M. Durante, T. Lecrevisse, J.-M. Rey, CEA DSM-IRFU-SACM, Gif-sur-Yvette, France
J. Fleiter, CERN, Geneva, Switzerland
M. Sorbi, G. Volpini, INFN Sezione di Milano LASA, Milano, Italy
P. Tixador, G2Elab/Institut Neel, CNRS/Grenoble-INP/UJF, Grenoble, France

*Abstract*

In the EuCARD project, we aim to construct a dipole magnet in YBCO producing 6 T in the background field of a 13 T $Nb_3Sn$ dipole FRESCA II. This paper reviews the quench analysis and protection of the YBCO coil. In addition, a recommendation for the protection system of the YBCO coil is presented.

## MAGNET SYSTEM

The YBCO dipole coil is a pre-accelerator magnet without beam tube. It consists of three stacked double pancake racetracks. A view of the coil is shown in Fig. 1. The magnet system is presented in detail in [1]. The magnet produces 6 T and it will be tested in the FRESCA II facility at CERN, providing a background field of 13 T [2].

The magnet will be wound from a cable made from two 12-mm-wide YBCO tapes manufactured by SuperPower. Additional copper stabilizer layers will be added to the cable. To reinforce the tape at its maximum operation field, 19 T, and high current density region, two 50 μm thick $CuBe_2$ layers are added to the cable. The cable is insulated with 30 μm thick Kapton insulation. A schematic view of the cable is shown in Fig. 2.

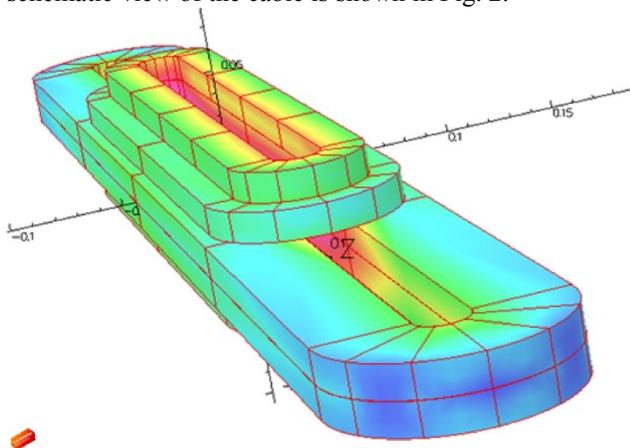

Figure 1: View of the YBCO racetrack insert.

In the coil, two of these cables are connected in parallel to reduce the coil inductance. The nominal operation current of the coil of 2800 A corresponds to an engineering current density of 252 A/mm$^2$. Critical current characteristics of the conductor were presented in [3]. Some characteristic values of the coil system, including the FRESCA II operating at bore magnetic field density of 13 T, are presented in Table 1. These parameters do not present the maximum operation conditions of FRESCA II at 1.8 K, but the design values for the combined operation with the YBCO insert at 4.2 K.

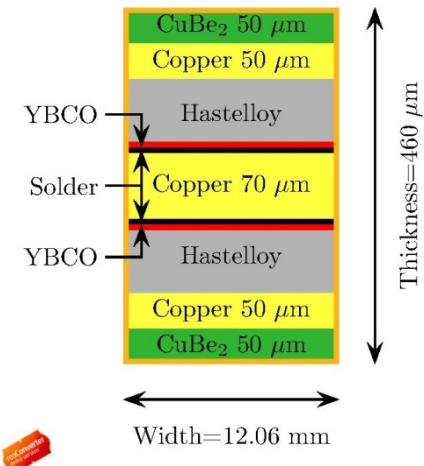

Figure 2: Schematic view of the cable for the YBCO racetrack insert.

Table 1: Parameters of the magnet system

| YBCO insert | | |
|---|---|---|
| Inductance | (mH) | 4 |
| Operation current | (A) | 2800 |
| Slef-energy | (kJ) | 15.7 |
| FrescaII | | |
| Inductance | (mH) | 98 |
| Operation current | (A) | 10500 |
| Self-energy | (kJ) | 5400 |
| Mutual inductance | (mH) | 9.3 |
| Total energy | (kJ) | 5690 |

## QUENCH MODELING AND PROTECTION SCHEME

Preliminary quench analysis and protection scheme of the magnet have been presented in [4]. Rest of the work will be presented later [5]. Here we consider how the temperature distribution evolves inside the YBCO coil when it quenches and how rapidly the current can be decreased to zero with an external dump resistor circuit. Due to the small energy stored in the insert, we considered only the external dump resistor circuits for the protection. We did not include the protection circuit into the finite element method based quench simulations.

Quenching an HTS magnet is difficult due to the high stability margin that HTS magnets typically have in large parts of the winding. There are, at least, two options to quench a magnet in simulations. First, one can simulate an additional heat input to some volume. This could, however, mean that in order to keep the volume small (length few mm of cable), one might need to increase the local temperature to as high as 100 K. This is not practical. Another option is to set a volume with reduced critical current to the coil. We used this option for quenching. The quench simulation time depends greatly on the size of this volume. We set this volume to the cable around the point where the minimum critical current was. To ignite the quench relatively fast, we set the length of this volume, part of the actual cable, to 10 cm and the critical current in this volume was fixed to 0 A. In [4] it was shown that the value of degraded critical current has negligible influence on the hot spot temperature at quench detection threshold voltage. Only the time from the beginning of the simulation to the quench detection changes considerably.

It is well known that in HTS coils the quench propagates very slowly and typically large part of the coil remains at operation temperature during a quench. Thus, we modelled only the top double pancake in the quench simulations. Different regions in the modelling domain are shown in Fig. 3.

We modelled part of the coil with insulation-tape-insulation structure and part of the coil was modelled as a homogeneous mixture of anisotropic material having effective material properties. In the upper coil, 11 of 35 turns were modelled with insulation-tape-insulation structure. In the lower coil 16 of 61 tapes were modelled in this way. Figure 4 shows the temperature distribution in that part of the coil where the temperature is higher than 8 K. Time instant is at the end of the simulation, i.e. maximum temperature is 400 K. As can be seen, the heat distributes only to a marginal part of the coil even at such a high hot spot temperature, thus demonstrating the feasibility of this approach.

In this simulation, the operation current was kept constant. Then, we can directly see what would be the temperature of the coil at given quench detection voltage. The simulation was terminated when the maximum temperature reached 400 K. Then the time was about 1 s. There were about 200 000 degrees of freedoms and the solution took about 50 hours. Custom-built finite element method quench software was used for the solution [4,6,7].

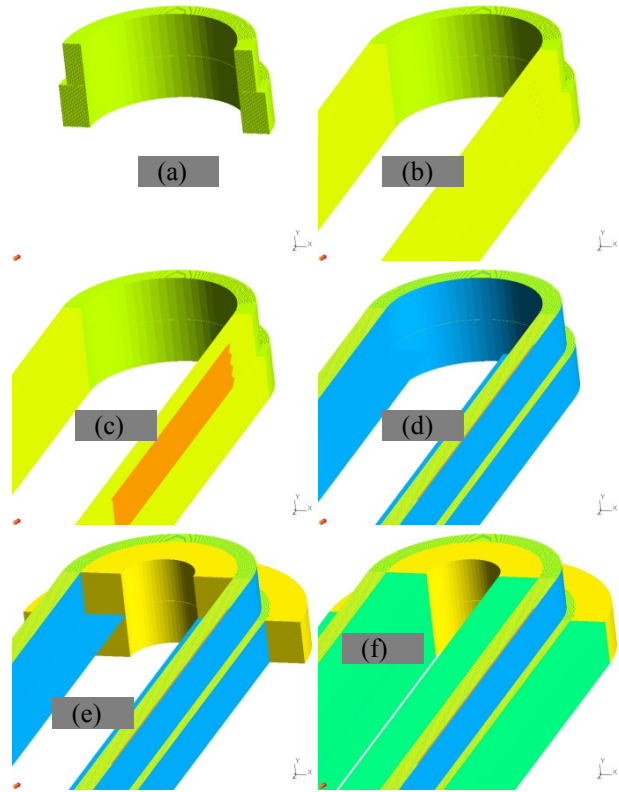

Figure 3: View of the different domains used in the quench simulations of the top double pancake shown in Fig. 1. One new domain type is added from subfigure to another. (a) tapes at the other end of the coil, (b) tapes at the straight parts, (c) volume with reduced critical current, (d) insulation, (e) other end modelled with effective homogeneous anisotropic material properties, (f) straight parts modelled with effective homogeneous anisotropic material properties.

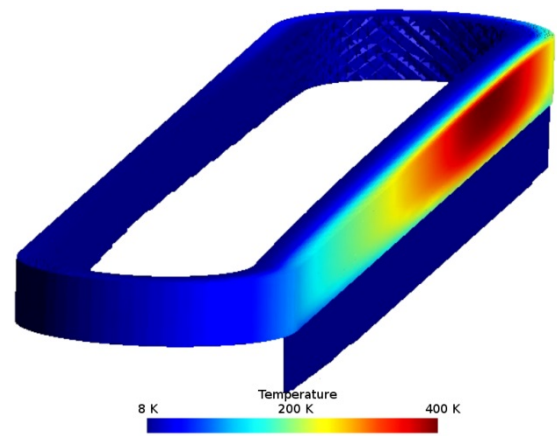

Figure 4: Temperature distribution at the end of the quench simulation.

Current sharing temperatures in the cross-section of the reduced domain and on top of the upper coil are shown in Fig. 5. These demonstrate the relatively large stability margin in the range 5-20 K.

the quench detection time instant. This is safe and certainly allows a rapid discharge before the coil is damaged due to overheating and insulation or YBCO layer melting. The thermal stresses due to the localized hot spot might pose other problems, but they are not considered here.

Two possible protection circuits of the insert are shown in Fig. 7. In the simpler one, the dump resistance is determined by maximum allowable terminal voltage of the magnet, 800 V. For the variable dump resistor, a gate-turn-off-thyristor (GTO) circuit can be used to adjust very rapidly the dump resistance in order to speed up the current decay. This method is effective for coils where the normal zone resistance grows very slowly and if quench heaters are not used. However, it complicates the protection circuit and as a new proposal introduces reliability risks before the circuit is actually built and tested.

Fig. 8 presents the operation of the protection circuit with time varying dump resistance. The steps for switch openings are determined by the quench analysis of FRESCA II [8]. Analysis is performed in such a way that the insert terminal voltage is kept at maximum in the range 800-1000 V.

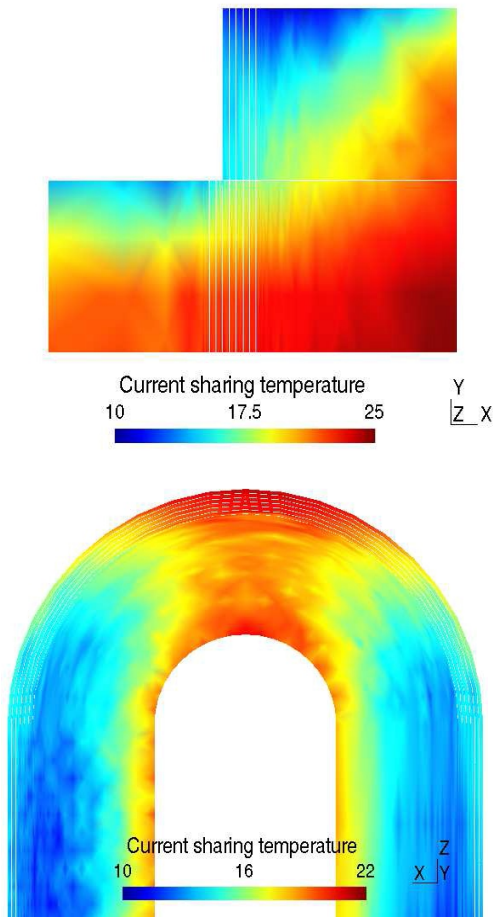

Figure 5: Current sharing temperature in the cross-section of the coil in the middle of the hot spot region (a) and on the top of the coil (b).

Especially in the lower coil, there are regions where a temperature increase to almost 25 K is required for current sharing temperature. In these simulations the redistribution of the current in one conductor during local heating was not considered, but the current density was always assumed to be homogeneously distributed in the cable and consequently heat generation was computed. It is of interest to develop modelling tools for this kind of modelling for the YBCO magnets of the future.

It is very unlikely that the middlemost coils quench, if the conductor is of good and homogeneous quality. Furthermore, the critical current decreases very slowly as a function of temperature and consequently heat generation increases slowly. This leads to slow normal zone resistance increase.

The hot spot temperature as a function of time and terminal voltage are shown in Fig. 6. The quench detection threshold voltage is expected to be 100 mV, corresponding to an hot spot temperature below 100 K at

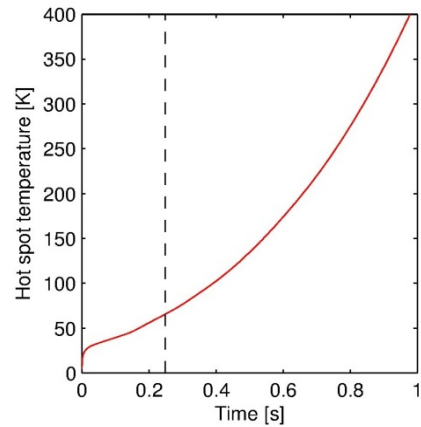
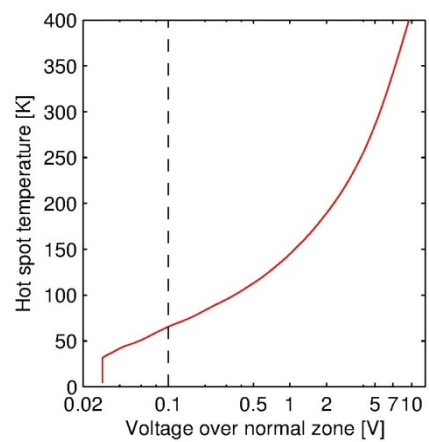

Figure 6: Hot spot temperature as a function of (a) time and (b) terminal voltage. Dashed lines present detection voltage threshold.

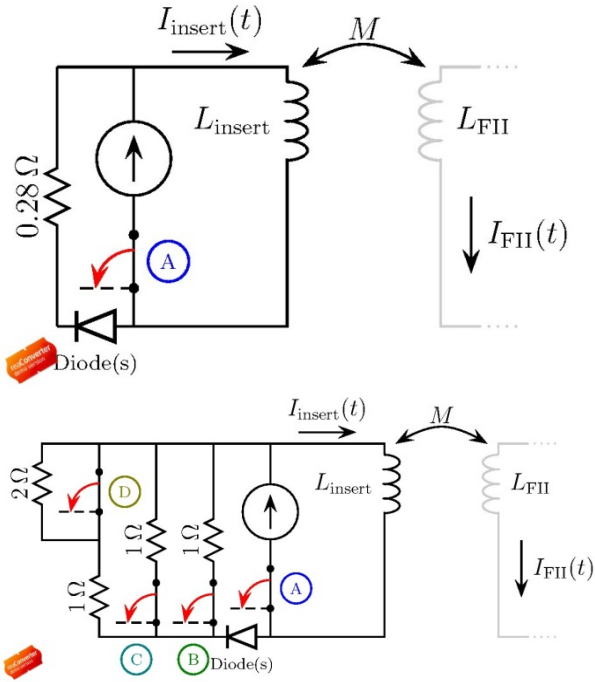

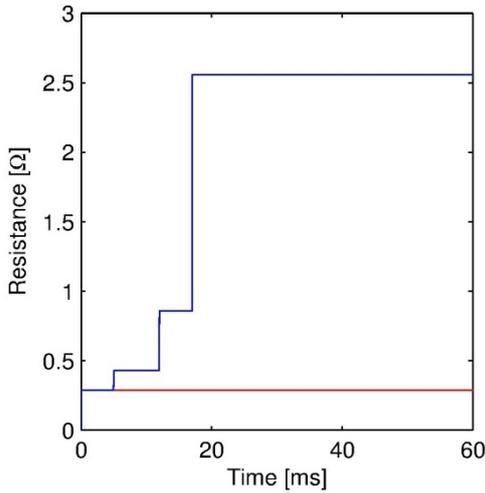

Figure 7: (a) basic protection circuit with a constant dump resistance. (b) protection circuit with varying dump resistance.

Figure 8: Effective resistance of the protection circuit during its operation.

Rapid insert discharge with the two possible protection circuits are illustrated in Fig. 9. With the time varying dump resistor, the current decay time can be reduced to 20 ms from 50 ms, still keeping the terminal voltage within acceptable limits at all times.

The variable dump resistor circuit allows the discharge of the insert in such a short time that the quench protection circuit of FRESCA II can be activated after the insert is in open circuit. This gives ultimate safety to the insert since no circulating currents are induced to it during the FRESCA II current decay. It is another question whether this is necessary or not because even with the constant dump resistor circuit no excess current is induced to the insert as shown in Fig. 10. Also in this case the insert terminal voltage remains within tolerable limits.

Finally, the open circuit voltage of insert during FRESCA II quench is shown in Fig. 11. This is clearly in safe limits. The rapid discharge of the insert might cause problems to the FRESCA II because of the induced current. This causes certain requirements to its power supply unit. It is to be explored whether this is tolerable or not and whether it is worth the risk to even consider this protection circuit. The simplest option would be to use the constant dump resistor circuit and in case of a quench, activate both the protection circuits simultaneously.

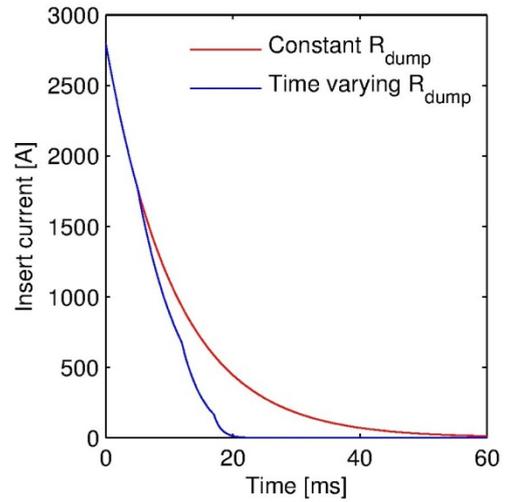

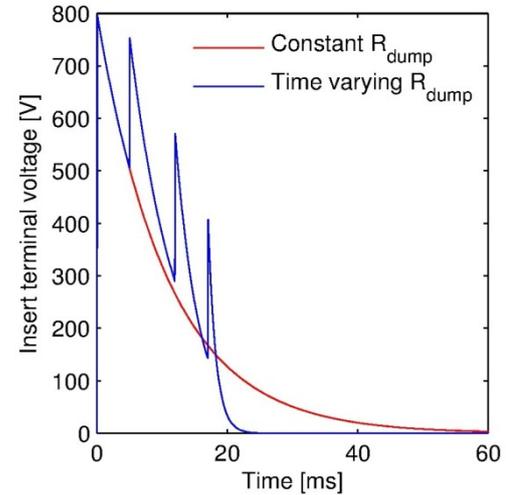

Figure 9: (a) insert current decay and (b) insert terminal voltage during insert rapid discharge.

## DISCUSSION

Such a small YBCO coil is not difficult to protect, at least, if the tape has homogeneous critical current and in a long-term use does not quench due to very small local damage. However, the critical current characteristic of the cable is doubtful. No widely used scaling law for the critical current of long unit length YBCO tape over wide range of magnetic field, its angle and temperature is available. Also, the tape properties vary from batch to batch depending on the doping of the coated conductor. Thus, the now used critical current hypersurface, including angular dependence of magnetic flux density, is not necessarily reliable.

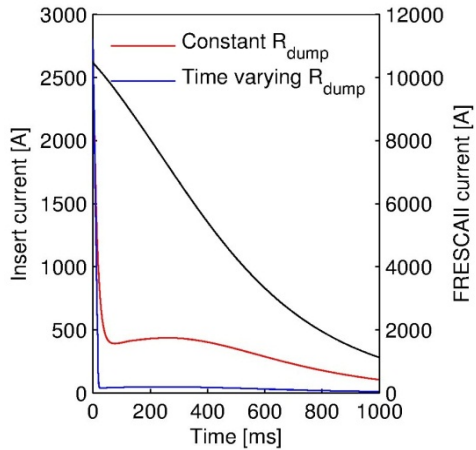

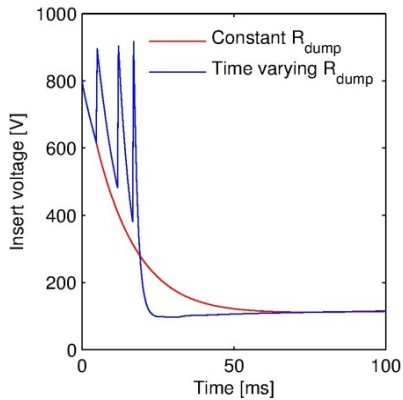

Figure 10: (a) FRESCA II [8] and insert current decay and (b) insert terminal voltage during FRESCA II quench.

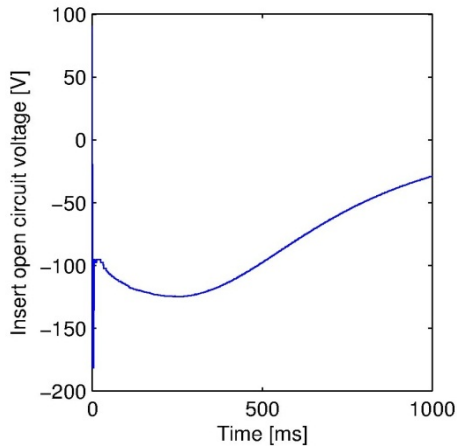

Figure 11: Insert open circuit voltage during FRESCA II quench.

Large HTS magnets can be very difficult to protect because quench heaters require high energies to ignite wide spread quench if compared to LTS counterparts. But before such detailed analysis can be performed, more fundamental research related to critical current hypersurface of coated conductors over wide range of parameters and heat generation at overcritical currents in YBCO is needed.

We also need to understand what kind of cable structures are beneficial for high current magnets and how current sharing occurs, because this has an important effect on the heat generation and thus normal zone propagation. Literature lacks this research especially in high field regime and at low temperatures. There is lot of quench related research of YBCO around 77 K but no conclusions should be drawn from these to 20 T region operation at 4.2 K.

In this work quench starting from the lowest critical current point was studied. Of course, the coil can quench also from the area where the field is parallel to the tape's wide surface, but this is highly unlikely, because the critical current of the cable at those regions is above 10 kA. Thus, a simulation at the low critical current region was only performed.

## RECOMMENDATION FOR THE PROTECTION OF THE COIL

Because the coil operates at such a high current at 4.2 K, the protection of the coil should be easier than that of coils operating around 20 K or even at 77 K, where the higher volumetric specific heat increases considerably the stability margin and consequently the hot spot temperature at the quench detection time. However, there are uncertainties in the modelling related to the critical current surface. The angular dependence is estimated and the critical current is just scaled from the 4 mm wide tape to the presented cable.

### Protection circuit

For simplicity, it is recommended to use only constant dump resistor for protection. The resistance should be sized according to the maximum allowed terminal voltage. If this value is 800 V, then the resistance is 0.28 Ω. However, to consider insert-outsert magnets of the future, research toward rapid discharge circuits could be started.

### Quench detection

At least in several earlier HTS coils the quench detection proved to be problematic due to the very slow normal zone propagation. Regardless of this, it is expected that a quench in this coil can be detected from the terminal voltage. However, for safety, it is recommended that, in addition to the terminal voltage, voltages over each coil are monitored and three differential signals are formed from these to separate the inductive voltages during magnet loading and discharge. Then, each of these voltages should include only resistive part, and this voltage could be used directly for detecting the quench. Noise level should be considerably below 100 mV. There is a possibility of symmetric simultaneous quenches, for example, in the topmost and bottommost pancakes, but then, the normal zone voltage grows twice as fast as in a single quench and therefore voltage allows quench detection.

The protection system of FRESCA II must also trigger the protection system of the insert and vice versa. Both systems can be launched in parallel, because it was shown that this causes no trouble to the insert. However, because the insert has much smaller energy than the FRESCA II and the coupling is good, high risk of damage to the insert and its power supply unit occurs, if the insert discharge switch does not open when the FRESCA II quench discharge begins. So there should be some redundancy for this aspect.

## CONCLUSIONS

Quench in dipole YBCO insert producing 6 T in the background field of 13 T was considered. The magnet is being built within the EuCARD project framework and will be tested in the bore of FRESCA II facility. We found out that this coil can be quenched safely. This is mainly due to the very low inductance and high current which allows quite rapid quench detection and very rapid current decay. Basic constant dump resistor protection circuit was introduced for the magnet. Also, a protection circuit with time dependent dump resistor was considered. A quench in this YBCO insert was concluded to be safe event and in case of FRESCA II quench, a simultaneous discharge of the magnets can be carried out.

## ACKNOWLEDGEMENTS

This work was supported by the European Commission under the FP7 Research Infrastructures project EuCARD, and is part of EuCARD Work Package 7: Superconducting High Field Magnets (HFM) for higher luminosities and energies.